# Direct evidence of the self-compression of injected electron-hole plasma in silicon

**Pavel Altukhov**[*,1] **and Evgenii Kuzminov**[1]

[1] A. F. Ioffe Physical-Technical Institute, Politekhnicheskaya Street 26, 194021 St. Petersburg, Russia

[*] Corresponding author: e-mail **pavel.altukhov@gmail.com**, Phone: +7 812 292 7344, Fax: +7 812 297 1017

A surface distribution of the electroluminescence intensity of silicon p-n light emitting diodes is obtained under space scanning experiments at room temperature. An emitting surface of the diodes, represented by a few small bright emitting dots and a weakly emitting area outside the dots, serves as a direct evidence of the self-compression of injected electron-hole plasma in silicon. The plasma self-compression explains concentration of injected carriers into one or a few strongly emitting plasma drops.

**1 Introduction** Nature of high quantum efficiency of silicon light emitting diodes at room temperature is an exciting problem of a large number of researches [1–12]. The internal quantum efficiency of electroluminescence of silicon diodes $\eta \sim 10^{-3} - 10^{-2}$ can be explained by the self-compression of electron-hole plasma into dense plasma drops or plasma flexes with the density of electron-hole pairs $n \sim 10^{17} - 10^{18}$ cm$^{-3}$ [1, 2]. A negative heat capacity of electron-hole plasma at the temperatures $T \geq 300$ K, concentration of the input diode power inside the plasma and weak diffusion of phonons at room temperatures represent the main physical reasons and conditions of the plasma condensation [2]. Under these conditions, generation of phonons by the plasma results in a local overheating of the lattice and a reduction of the semiconductor energy gap inside the plasma. In such a way the self-organized potential well, attracting injected electrons and holes, is created [2]. At the negative heat capacity of the plasma, the plasma self-compression is accompanied by a decrease of plasma energy, observed as a low energy shift of the high energy edge of a recombination radiation line in electroluminescence spectra with increasing diode current [2]. Here we represent direct evidence of the plasma self-compression in silicon p-n light emitting diodes at room temperatures, using results of a surface scanning of the electroluminescence intensity.

**2 Results and discussion** Silicon p-n light emitting diodes were fabricated on n-type silicon substrates with a phosphorus concentration of about $10^{14}$ cm$^{-3}$. The p-n junction was formed by a surface layer, highly doped by boron under thermal diffusion, with a boron concentration of about $5\times10^{19}$ cm$^{-3}$. The diodes with circular and rectangular shapes and sizes equal to (1÷10) mm were used in the experiments. The emitting surface of the diode with rectangular shape is shown in Figure 1.

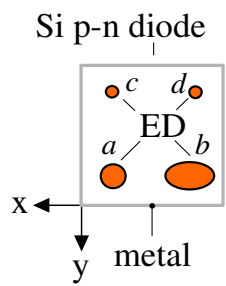

**Figure 1** The emitting dots (ED, *a*, *b*, *c*, *d*) of the silicon p-n light emitting diode before the switching of the EL intensity distribution. The diode area is $S = 4\times4$ mm$^2$.

The thickness of the silicon layer, highly doped by boron, is about 1 micron. The Al strip metal layer with the thickness of the layer about 1 micron and the layer width about 0.3 mm was made on the emitting surface of the diodes (Figure 1). A surface layer, highly doped by phosphorus under thermal diffusion, with a phosphorus concentration of about $5\times10^{19}$ cm$^{-3}$ serves as a metal contact on the back side of the diodes. A Ni cover metal layer with a thickness of the layer of about 1 micron was made on the back side of the diodes. The thickness of the silicon substrates is 0.35 mm. The gate voltage of the diodes $V_g$ is applied between the top Al strip and the bottom Ni layer.

The image of the emitting surface of the diode was focused by lenses on an input slit of a monochromator with the width and height of the slit equal to 0.15 mm. The recombination radiation was detected by a cooled photomultiplier using a photon counting system at a photon energy equal to 1.13 eV. The surface scanning of the electroluminescence intensity was produced by a step movement of the image along the slit with the space step equal to 0.1 mm along the x and y direction. The profiles of the surface distribution of electroluminescence (EL) intensity along the x and y directions are shown in Figure 2.

The image of the emitting surface is represented by four emitting dots (Figures 1 and 2), located at the corners of the diode, and only two or one of them (*a*, *b*) give a strong electroluminescence. The surface distribution of the EL intensity appears to be bistable. The profile 2x of the EL intensity with the maximum at the dot *b* switches to the profile 1x with the maximum at the dot *a*. This is accompanied by a switching of the gate voltage of the diode from 2.29 V to 2.06 V at the diode current equal to 0.2 A. The surface profile width of the dot emission is defined by light propagation along silicon plates and depends on quality of the silicon surface. For surfaces of high quality the profile width is about 0.7 mm. The minimum profile width, equal to 0.3 mm, is achieved for surfaces with a large scale roughness.

The voltage-current dependences of these diodes show a monotone increase of the voltage with increasing charge current, and we attribute the observed light emission to electron-hole plasma drops [2] with the maximum size equal to 0.3 mm at the maximum diode current equal to 0.2 A. The value of the internal quantum efficiency of the diodes is $\eta \approx 0.03$, and the corresponding minimum electron-hole density in the drops is $n \sim 10^{17}$ cm$^{-3}$ [2]. For calculations of the plasma density we use the value of the recombination radiation coefficient $A = 3\times10^{-15}$ cm$^3$ s$^{-1}$ [2], obtained by the Roosbroeck and Shockley calculation procedure [13]. A contribution of excitons to the low energy edge of the emission line could enhance the value of $A$ by 1.5 times [14]. But excitons are screened at high plasma densities, and the exciton contribution is assumed to be absent.

The conclusion about existence of injected electrons and holes in the form of dense plasma drops is supported by the electroluminescence spectra of the diodes (Figure 2, spectra 3, 3'). The strongly shifted to the low energy side of the spectrum emission lines of the diodes in comparison with the calculated emission line of rare electron-hole gas (Figure 2, spectrum 3") give evidence of an essential local overheating of the lattice inside the plasma drops [2].

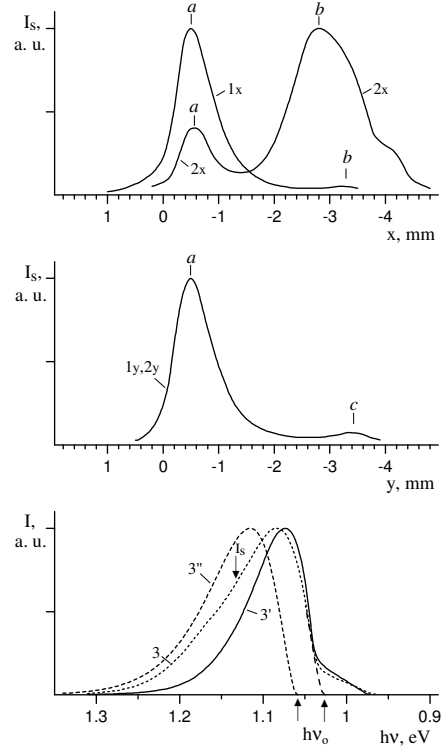

**Figure 2** Surface distribution of the EL intensity $I_S$ (1x, 1y, 2x, 2y) at $h\nu = 1.13$ eV for the emitting dots (ED, *a*, *b*, *c*) and EL spectra (3, 3', 3") of the silicon p-n diode.
x is the distance from the left edge of the diode in the x direction; y is the distance from the lower edge of the diode in the y direction (Figure 1).
The diode current is $J^o = 0.2$ A,
the diode temperature is $T = 320$ K,
the temperature of the e-h plasma drops is $T_g = 440$ K.
1x, 1y: after the switching
of the EL intensity maximum from the dot *b* to the dot *a*;
2x, 2y: before the switching.
y = - 0.5 mm for 1x, 2x;
x = - 0.5 mm for 1y, 2y.
The diode voltage $V_g$ is 2.29 V before the switching and 2.06 V after the switching.
3, 3': EL spectra after the switching.
3: at the maximum of the dot *a*
for y = - 0.5 mm, x = - 0.5 mm.
3': for y = - 0.5 mm, x = - 2.2 mm.
3": theory for electron-hole gas at $T = 320$ K.
The arrow $I_S$ at the spectrum 3 indicates the photon energy for the space scanning experiments. The arrows $h\nu_o$ indicate the position of the low energy edge of the TO emission lines [2].

The spectral position of the low energy edge of the emission lines $h\nu_o$ (Figure 2) gives the value of the local temperature of the electron-hole plasma drops [2]. The local temperature of plasma drops is $T_g = 440$ K at the diode current $J^o = 0.2$ A, and the diode temperature outside the drops is $T = 320$ K. Nevertheless, this significant local overheating is not sufficient to explain all features of the plasma self-compression, especially, a significant difference between the carrier densities in the drops and outside the drops. We assume that the self-focusing of the injection current in the region of the plasma drops is an additional physical process, necessary for condensation of injected carriers into plasma drops. This self-focusing of the injection current can accompany the plasma self-compression because of the reduction of the semiconductor energy gap in the region of plasma drops, resulting in a local enhancement of the injection current. Sites of concentration of the injection current can be defined by the potential inhomogeneities. Note, that for some diodes one or two plasma drops are located at center of an edge of a diode.

The self-focusing of injection current explains a low density of injected carriers outside the drops with a low probability of radiative recombination $An$ and a very weak light emission from injected carriers outside the drops. The emission line 3' in the EL spectrum (Figure 2) is attributed to the emitting dot $a$. Light propagation from the emitting dot along the silicon plate results in absorption of light in the high energy part of the spectrum and the corresponding narrowing of the emission line under measurements of the spectra at large distances from the emitting drop. A uniform surface distribution of the EL intensity of p-n diodes was obtained with use of integrated emission with a large contribution of the low energy part of the emission line [11]. This part of the emission propagates with a low absorption for a long distance, resulting in the observed [11] low resolution of space scanning experiments.

Our model of the plasma self-compression seems to be correct for the explanation of presented experimental results. Resent experiments [12] show, that the model of dislocation loops [3, 4], concentrating injected carriers near a p-n junction, is not available for explanation of the high quantum efficiency of silicon diodes. Defects, introduced by boron implantation and used as a cause of the high quantum efficiency of silicon diodes [9, 10], are absent in our diodes.

**3 Conclusion** The results of our experiments can not be explained by the simple Shockley diffusion model of the p-n junction current, considered in Ref [8]. The concentration of the electroluminescence intensity in small emitting dots on the diode surface, defined by the concentration of injected carriers in dense plasma drops, and the low energy shift of the emission line serve as direct evidence of the self-compression of injected carriers in silicon p-n light emitting diodes. The introduced self-focusing of injection current can represent an essential feature of the phenomenon, necessary for the realization of high-contrast images of the surface distribution of light emission.